\documentclass[twocolumn,showpacs,preprintnumbers,amsmath,amssymb]{revtex4}
\usepackage{graphicx}% Include figure files
\usepackage{dcolumn}% Align table columns on decimal point
\usepackage{bm}% bold math
\usepackage{color}

\begin{document}

\preprint{PRL}

\title{Li diffusion in Li$_x$CoO$_2$ probed by muon-spin spectroscopy
}

\author{Jun~Sugiyama$^1$}
 \email{e0589@mosk.tytlabs.co.jp}
\author{Kazuhiko~Mukai$^1$}
\author{Yutaka~Ikedo$^1$}
 \altaffiliation[Present address: ]{Muon Science Laboratory, 
Institute of Materials Structure Science, 
High Energy Accelerator Research Organization, Tsukuba}, 
\author{Hiroshi~Nozaki$^1$}
\author{Martin~M\aa{}nsson$^2$}
\author{Isao~Watanabe$^3$}

\affiliation{%
$^1$Toyota Central Research and Development Laboratories~Inc.,
Nagakute, Aichi 480-1192 Japan
}%

\affiliation{%
$^2$Laboratory for Neutron Scattering, ETH Z$\ddot{\rm u}$rich
and Paul Scherrer Institut, CH-5232 Villigen PSI, Switzerland
}%

\affiliation{$^3$Muon Science Laboratory, RIKEN, 2-1 Hirosawa,
Wako, Saitama 351-0198, Japan
}%

\date{\today}% It is always \today, today,

\begin{abstract}
The diffusion coefficient of Li$^+$ ions ($D_{\rm Li}$)
in the battery material Li$_x$CoO$_2$ has been investigated
by %means of 
muon-spin relaxation ($\mu^+$SR). 
%because $D_{\rm Li}$ for positive electrode materials 
%has not been determined correctly so far. 
Based on %performing  
the experiments in 
zero-field and weak longitudinal-fields at temperatures up to 400 K, 
we determined the fluctuation rate ($\nu$) of the fields 
on the muons due to their interaction with the nuclear moments.
Combined with susceptibility data and electrostatic potential calculations,
clear Li$^+$ ion diffusion was detected above $\sim150~$K.
The $D_{\rm Li}$ estimated from $\nu$ was in very good agreement
with predictions from first-principles calculations,
and we present the $\mu^+$SR technique as an % novel and 
optimal probe to detect $D_{\rm Li}$
%of unique usefulness 
for materials containing magnetic ions.
\end{abstract}

\maketitle

In spite of a long research history on lithium insertion materials for Li-ion batteries
\cite{Ohzuku}, e.g., LiCoO$_2$, LiNiO$_2$, and LiMn$_2$O$_4$,
one of their most important intrinsic physical properties, 
the Li$^+$ ions diffusion coefficient ($D_{\rm Li}$),
has not yet been determined with any reliability.
%Many researchers in condensed matter physics 
%along with people in the muon-spin rotation and relaxation ($\mu^+$SR)community, 
%would imagine that 
Although Li-NMR is, in general, a powerful technique to measure $D_{\rm Li}$ 
for non-magnetic materials, 
it is particularly difficult to evaluate $D_{\rm Li}$ for materials containing magnetic ions, 
because the magnetic ions induce additional pathways
for the spin-lattice relaxation rate ($1/T_1$),
resulting in huge $1/T_1$ compared with that expected for only the diffusive motion of Li ions.

Such difficulty was clearly evident in the $1/T_1(T)$ curve
for LiCoO$_2$ and LiNiO$_2$ \cite{Tomeno,Nakamura}, and, 
for that reason, $D_{\rm Li}$ was instead estimated from the Li-NMR line width \cite{Nakamura_2}.
However, since the line width, i.e., the spin-spin relaxation rate ($1/T_2$)
is also affected by the magnetic ions,
the $D_{\rm Li}$ obtained by Li-NMR for LiCoO$_2$
($=1\times10^{-14}~$cm$^2$s$^{-1}$ at 400~K) is approximately four orders of magnitude smaller 
than predicted by first-principles calculations \cite{Ven}. 
Since lithium insertion materials always include transition metal ions, 
in order to maintain charge neutrality
during the extraction and/or insertion of Li$^+$ ions, 
it is consequently very difficult to determine $D_{\rm Li}$
for these compounds unambiguously by Li-NMR.
%Here, $D_{\rm Li}$ is naturally one of the most predominant parameters
%to govern the charge/discharge rate of the Li-ion battery.

On the other hand,
the chemical diffusion coefficient ($D_{\rm Li}^{\rm chem}$),
which is measured under a potential gradient,
has also been determined by electrochemical measurements.
Note that the relationship between $D_{\rm Li}$ and $D_{\rm Li}^{\rm chem}$ is given by
$D_{\rm Li}^{\rm chem}=\Theta D_{\rm Li}$,
where $\Theta$ is a thermodynamic factor.
The magnitude of $D_{\rm Li}^{\rm chem}$ is, however, known to be very sensitive to
the measurement system,
e.g., the electrolyte as well as the compositions of the positive and negative electrodes.
As a result, the reported $D_{\rm Li}^{\rm chem}$ for Li$_x$CoO$_2$
ranges from $4\times10^{-8}$ to $10^{-10}$~cm$^2$s$^{-1}$ for powder samples
\cite{Mizushima_2,Yao,Pyun,Dokko1,Dokko2} and
from $2.5\times10^{-11}$ to $2\times10^{-13}$~cm$^2$s$^{-1}$ for thin films 
\cite{Striebel,Rho,Xia,Tang} 
at ambient $T$.
In order to profoundly understand the physics behind the operation principle
of battery materials,
it is imperative to have a reliable probe to measure $D_{\rm Li}$ for all the components
of the battery as a function of both Li content as well as $T$.
This is at present a key issue for further development of Li-ion batteries, 
and in particular for future fabrication of solid-state batteries.

In contrast to NMR,
the effect of localized moments in a paramagnetic (PM) state
on the $\mu^+$SR signal
is very limited at high $T$,
because the PM fluctuation is usually too fast to be {\it visible} by $\mu^+$SR.
Furthermore, although $\mu^+$SR is very sensitive to the local magnetic environment,
whether it is due to electronic or nuclear spins, 
an electronic contribution is, in principle, distinguishable from a nuclear contribution 
by weak longitudinal field measurements. 
In addition, since the $\mu^+$SR timescale is different from the NMR's one,  
$\mu^+$SR is expected to provide unique information on nuclear magnetic fields. 
Indeed, $\mu^+$SR studies of Li$_x$CoO$_2$ at low $T$
indicate that $\mu^+$s {\it feel} a nuclear magnetic field
caused by Li and $^{59}$Co even at 1.8~K \cite{Jun_1,Mukai}.
This offers a possibility to determine $D_{\rm Li}$ of Li$_x$CoO$_2$ by $\mu^+$SR,
if $\mu^+$s locate in the vicinity of the O$^{2-}$ ion
and make a stable $\mu^+$-O$^{2-}$ bond in the Li$_x$CoO$_2$ lattice.
Here, we report our initial work on Li$_x$CoO$_2$ ($x$=0.73 and 0.53) 
to estimate their $D_{\rm Li}$ and 
establish $\mu^+$SR as a
novel tool to probe Li-ion diffusion. 

%\section{\label{sec:Intro}Experiment}

A powder sample of LiCoO$_2$ was prepared
at Osaka City University by a solid-state reaction technique
using reagent grade LiOH$\cdot$H$_2$O
and CoCO$_3$ powders as starting materials.
A mixture of the two powders
was heated at 900$^{\rm o}$C for 12~h in air.
Powder X-ray diffraction (XRD) analysis showed that
the LiCoO$_2$ sample was single phase with a rhombohedral system
of space group $R\overline{3}m$
($a_{\rm H}=0.2814$ nm and $c_{\rm H}=1.4049$~nm
in hexagonal setting).
The Li-deficient samples
were prepared by an electrochemical reaction
using Li$\mid$LiPF$_6$-ethylene carbonate-diethyl carbonate$\mid$LiCoO$_2$ cells.
The LiCoO$_2$ powder was pressed into a disc with 15~mm diameter and 0.4~mm thickness,
and the disc was then used as a positive electrode.
The Li$_x$CoO$_2$ disk was removed from the cell in a glove-box and packed into a sealed powder cell 
just before the $\mu^+$SR measurement.
%In order to assess possible changes in the Li-deficient samples
%during the $\mu^+$SR measurement,
%each sample was afterwards returned to the original cell
%to check their voltage (vs. a Li electrode).
%No significant change in the voltage was observed before and after the measurements.
Their structures were subsequently confirmed by powder XRD,
and, finally, their compositions were checked by an inductively coupled plasma 
atomic emission spectral analysis.
The above procedure is essentially the same as that of our previous $\mu^+$SR work
on Li$_x$CoO$_2$ \cite{Mukai} and Li$_x$NiO$_2$ \cite{Jun_2}.

The $\mu^+$SR spectra were measured
at the {\bf ARGUS} surface muon beam line of
the RIKEN-RAL Muon Facility at ISIS in the UK 
using a liquid-He flow type cryostat in the $T$ range between 10 and 400~K.
%the {\sf $\pi$A} surface muon beam line at KEK,
%the {\sf $\pi$M3-GPS} surface muon beam line at PSI, and
%the {\sf M20} surface muon beam line at TRIUMF.
The experimental techniques
were described elsewhere \cite{Kalvius}.
$\chi$ was measured
using a %superconducting quantum interference device
SQUID magnetometer
(MPMS, Quantum Design)
in the $T$ range between 5 and 200~K
under a magnetic field of $H$~=~100~Oe.

%\section{\label{sec:Intro}Results and Discussion}
%\subsection{\label{sec:Intro}overall nature}
\begin{figure}[t]
  \begin{center}
    \includegraphics[keepaspectratio=true,width=60 mm]{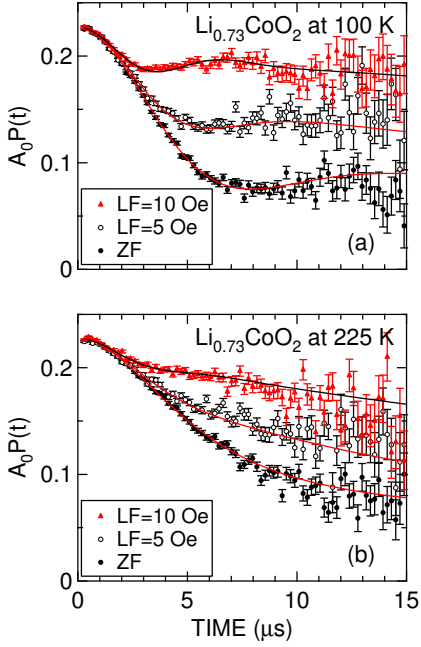}
  \end{center}
  \caption{(Color online)
  ZF- and two LF-$\mu^+$SR spectra for Li$_{0.73}$CoO$_2$ measured at
  (a) 100~K and
  (b) 225~K.
  The magnitude of LF was 5 and 10~Oe.
  Solid lines represent the fit result using Eq.~(\ref{eq:DKT}).
  }
  \label{fig:spectrum}
\end{figure}
Figure~\ref{fig:spectrum} shows the zero field (ZF-) and
longitudinal field (LF-) $\mu^+$SR spectrum
for the Li$_{0.73}$CoO$_2$ sample
obtained at 100 and 225~K.
At 100~K,
the ZF-spectrum exhibits a typical Kubo-Toyabe (KT) behavior
with a minimum at $t\sim6~\mu$s,
meaning that the implanted muons {\it see} the internal magnetic field ($H_{\rm int}$)
due to the nuclear magnetic moments of $^{7}$Li, $^{6}$Li and $^{59}$Co.
The applied LF clearly reduces the relaxation rate, i.e.,
the time slope,
by {\it decoupling} $H_{\rm int}$.
Although the ZF-spectrum still shows KT behavior at 225~K,
the relaxation rate is smaller than at 100~K.

\begin{figure}[t]
  \begin{center}
    \includegraphics[keepaspectratio=true,width=\columnwidth]{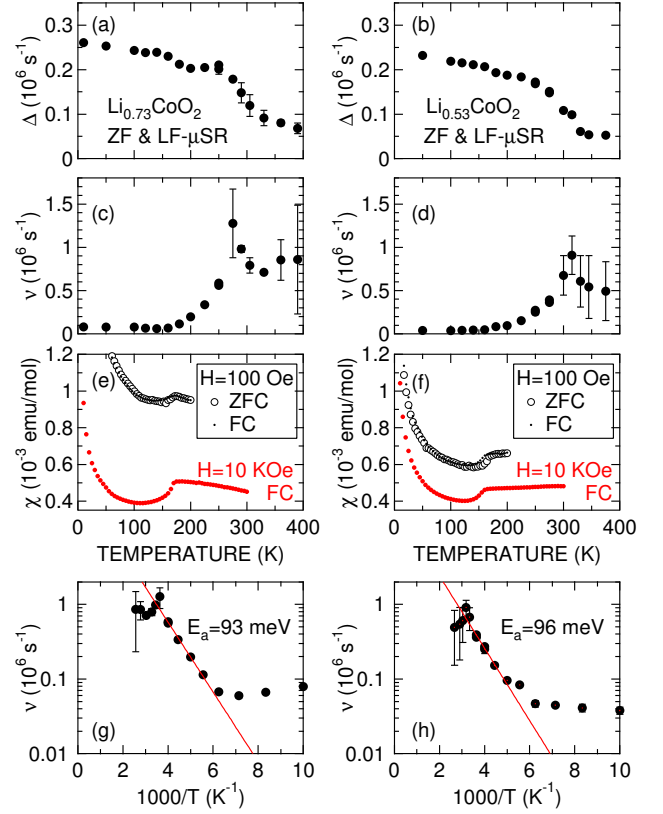}
  \end{center}
  \caption{(Color online) 
  For Li$_{0.73}$CoO$_2$ and Li$_{0.53}$CoO$_2$ respectively, we show
  $T$ dependences of
  (a, b) field distribution width ($\Delta$), 
  (c, d) field fluctuation rate ($\nu$), 
  (e, f) susceptibility ($\chi$), and
  (g, h) the relationship between log($\nu$) and $1/T$. 
  $\Delta$ and $\nu$ were obtained by fitting both ZF- and LF-spectra 
  using Eq.~(\ref{eq:DKT}). 
  $\chi$ was measured in both field cooling ($FC$) and zero field cooling ($ZFC$) mode
  with $H$=100~Oe. 
  In (e, f), the $\chi$ data \cite{Mukai} measured in $FC$ mode with $H=10~$kOe 
  for Li$_{0.75}$CoO$_2$ and Li$_{0.52}$CoO$_2$ 
  were also plotted for comparison. 
%  $T$ dependences of
%  (a) [(e)] the field distribution width ($\Delta$),
%  (b) [(f)] the field fluctuation rate ($\nu$), and
%  (c) [(g)] susceptibility ($\chi$) and
%  the relationship between log($\nu$) and $1/T$ (d) [(h)]
%  for Li$_{0.73}$CoO$_2$ [Li$_{0.53}$CoO$_2$].
%  $\chi$ was measured in both field cooling ($FC$) and zero field cooling ($ZFC$) mode
%  with $H$=100~Oe. 
The straight lines in (g) and (h) show the activated diffusive behaviour discussed in the text.
  }
  \label{fig:DN}
\end{figure}

In order to estimate the KT parameters precisely,
the ZF- and two LF-spectra were fitted simultaneously
by a combination of
a dynamic Gaussian KT function 
[$G^{\rm DGKT}(\Delta, \nu, t, H_{\rm LF})$]
and an offset signal
from the fraction of muons stopped mainly in the sample holder,
which is made of high-purity aluminum;
\begin{eqnarray}
 A_0\,P_{\rm LF}(t) &=&
  A_{\rm KT} G^{\rm DGKT}(\Delta, \nu, t, H_{\rm LF})+ A_{\rm BG}
\label{eq:DKT}
\end{eqnarray}
where $A_0$ is the empirical maximum muon decay asymmetry,	
$A_{\rm KT}$ and $A_{\rm BG}$ are the
asymmetries associated with the two signals.
$\Delta$ is the static width of the local field distribution
at the disordered sites,
and $\nu$ is the field fluctuation rate.
When $\nu=0$ and $H_{\rm LF}=0$, $G^{\rm DGKT}(t,\Delta,\nu, H_{\rm LF})$ is the
static Gaussian KT function $G_{zz}^{\rm KT}(t,\Delta)$ in ZF. 
At first, we fitted all the ZF-spectra using common $A_{\rm KT}$ and $A_{\rm BG}$ in Eq.~(\ref{eq:DKT}).  
The ``global fit" provided that 
$A_{\rm KT}=0.164056\pm0.000011\ (0.16889\pm0.00018)$ and 
$A_{\rm BG}=0.06350\pm0.00002\ (0.0692\pm0.0002)$ 
for Li$_{0.73}$CoO$_2$ (Li$_{0.53}$CoO$_2$). 
Then, using the obtained $A_{\rm KT}$ and $A_{\rm BG}$, 
one ZF- and two LF-spectra were global-fitted 
using common $\Delta$ and $\nu$ at each $T$.

%given by:
%
%\begin{eqnarray}
% &~& G_{zz}^{\rm KT}(t,\Delta)~=~\frac{1}{3}
% ~+~\frac{2}{3}(1 - \Delta^2 t^2)\exp(- \frac{\Delta^2 t^2}{2}).
%\label{eq:GKT}
%\end{eqnarray}
%

%
\begin{figure}[t]
  \begin{center}
    \includegraphics[keepaspectratio=true,width=\columnwidth]{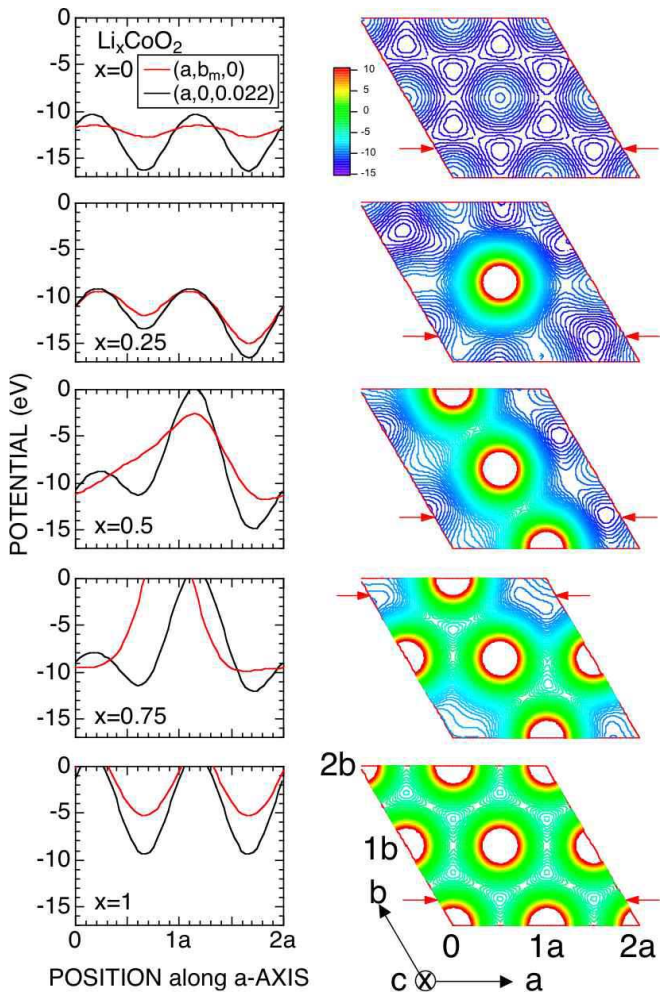}
  \end{center}
  \caption{(Color online)
  The variation of electrostatic potential ($\phi_{\rm E}$) along the $a$-axis
  at $b=b_m$ and $c=0$ (on the Li plane)
  and $b=0$ and $c=0.022$ (1~\AA\ away from the O$^{2-}$ ions)
  in the hexagonal lattice of Li$_x$CoO$_2$
  with $x=0,\ 0.25,\ 0.5,\ 0.75,$ and 1 (from top to bottom). 
  Here, $b_m$ denotes the $b$, at which electrostatic potential exhibits a minimum.
  The right panels show the corresponding distribution of
  $\phi_{\rm E}$ in the Li plane. 
  Arrows on the $b$-axis in the right panels represents $b_m$.
  }
  \label{fig:Potential}
\end{figure}

Figure~\ref{fig:DN} shows the $T$ dependencies of both $\Delta$ and $\nu$
for the two samples together with $\chi$ measured in a ~100~Oe magnetic field. 
For Li$_{0.73}$CoO$_2$, $\Delta$ is almost constant for $5\leq{}T\leq{}250~$K, 
indicating that the $\mu^+$ are most probably stable 
in the crystal lattice until $\sim300~$K. 
$\Delta$ then rapidly decreases,
but levels off again for $T\geq$~325~K.
%For Li$_{0.73}$CoO$_2$, as $T$ increases from 5~K,
%$\Delta$ decreases slowly up to 150~K,
%and then decreases faster with further increasing $T$ up to 200~K,
%then levels off to a constant value up to 250~K,
%decreases more rapidly up to 300~K,
%and finally decreases slowly again at $T$ above 325~K.
The $\nu(T)$ curve is almost $T$-independent up to 150~K,
starts to increase at $\sim150~$K,
and exhibits a maximum at 275~K.
Above 275~K, $\nu$ decreases to $0.7\times10^6~$s$^{-1}$ at $\sim325~$K,
and finally becomes almost $T$-independent above 350~K.
%if we consider a large estimation error.
The increase in $\nu$ between 150 and 275~K is well explained
by a thermal activation process [Figs.~\ref{fig:DN}(g) and \ref{fig:DN}(h)], 
which signals the onset of diffusive motion of
either Li$^+$ or $\mu^+$ above 150~K.
The clear decrease in $\Delta$ at $\sim300~$K also suggests
an additional diffusion of Li$^+$ or $\mu^+$.

The $\chi(T)$ curve exhibits a small anomaly
around 150~K with a thermal hysteresis of $\sim10~$K,
while there is no indication of any magnetic %transitions by $\mu^+$SR in this $T$ range 
anomalies in the $T$ range between 200 and 300~K [Figs.~\ref{fig:DN}(e) and \ref{fig:DN}(f)]
\cite{Jun_1,Mukai}. 
This suggests that the change in the $\mu^+$SR parameters around 150~K 
is caused by an intrinsic change in Li$_x$CoO$_2$,  
but the change around 300~K is visible only by $\mu^+$SR.  
%Indeed, the $\Delta(T)$ curve shows a small decrease with $T$ around 150~K, 
%revealing a slight change in a local magnetic environment in this $T$ range. 
The increase in $\nu$ above 150~K is, thus, most unlikely due to $\mu^+$ diffusion
but it is rather due to Li$^+$ diffusion, 
i.e., either a freezing of the Li$^+$ motion or 
an order-disorder transition of the Li$^+$ ions occurs below around 150~K.
This is also supported by a recent $^7$Li-NMR experiment \cite{Nakamura_3}, 
in which the NMR line width -vs.-$T$ curve for Li$_{0.6}$CoO$_2$ 
exhibits a step-like decrease with $T$ around 150~K 
by motional narrowing due to Li$^+$ diffusion. 
Since such diffusion naturally increases a local structural symmetry, 
it is reasonable that $\Delta$ slightly decreases with $T$ around 150~K. 
On the other hand, 
both Li$^+$ and $\mu^+$ are inferred to be diffusing above 300~K,
resulting in the large decrease in $\Delta$
caused by motional narrowing.
Actually, because $\Delta\leq0.1\nu$ above 300~K,
Eq.~(\ref{eq:DKT}) is roughly equivalent to an exponential relaxation function
[$\exp(-\lambda t)$], and it is 
%It is, hence, 
difficult to estimate $\Delta$ and $\nu$ precisely at high $T$. 

The result for Li$_{0.53}$CoO$_2$ sample is very similar to that of Li$_{0.73}$CoO$_2$,
although $\Delta_{T\rightarrow0}$(Li$_{0.53}$CoO$_2)
<\Delta_{T\rightarrow0}$(Li$_{0.73}$CoO$_2$)
due to the decrease in the number density of Li$^+$ ions,
as reported previously \cite{Jun_1,Mukai}.
Also, the magnitude of $\nu$ of Li$_{0.53}$CoO$_2$ is smaller
than $\nu$ of Li$_{0.73}$CoO$_2$ in the whole $T$ range measured,
but the $\nu(T)$ curve for both samples show a
clear increase with $T$ above 150~K and a maximum around 300~K.

In order to predict the muon site(s)
and to confirm the reliability of the above assumption
that Li$^+$ ions diffuse above 150~K
whereas $\mu^+$ diffuse only above 300~K,
we performed electrostatic potential ($\phi_{\rm E}$) calculations
for the Li$_x$CoO$_2$ lattice using a point-charge model
and the program DipElec \cite{Kojima}.
As seen in Fig.~\ref{fig:Potential},
the site in the vicinity of the O$^{2-}$ ions is
more stable for $\mu^+$ than the site in the Li plane
for the whole $x$ range between 1 and 0.
This means that
$\mu^+$'s are bound to the O$^{2-}$ ions 
so as to make a stable $\mu^+$-O$^{2-}$ bond in Li$_x$CoO$_2$. 
This is a common situation in oxides, 
as for example in the case for the high-$T_c$ cuprates \cite{Adams}. 
Since $\mu^+$s are assigned as an ideal point charge, 
such $\mu^+$-O$^{2-}$ bond should be purely ionic. 
In fact, dipole field calculations for the site in the vicinity of the O$^{2-}$ ions 
provide that 
$\Delta_{\rm calc}=0.43\times10^6$~s$^{-1}\ (0.35\times10^6$~s$^{-1}$) 
for Li$_x$CoO$_2$ with $x=3/4\ (1/2)$. 
Furthermore, $\Delta_{\rm calc}$ is found to be comparable to $\Delta$ measured at low $T$  
in the whole $x$ range for Li$_x$CoO$_2$ \cite{Mukai}, 
if we consider the reduction of $\Delta$ 
by the electric field gradient effect on the nuclear moments with $I\geq1$ \cite{Hayano,Kaiser}.   
This suggests that the point-charge model is acceptable for determining the muon site(s)
in Li$_x$CoO$_2$.
As a result, it is clarified that,
as $T$ increases from 5~K,
the Li$^+$ ions start to diffuse above 150~K ($=T_d^{\rm Li}$) and then
the $\mu^+$ diffuse above 300~K ($=T_d^{\mu}$),
in spite of the mass difference between $\mu^+$ and Li$^+$
($m_{\rm Li^+}/m_{\mu^+}\sim63$) because the muons form a hydrogen-like bond with oxygen.
%Note that the difference of the potential minimum ($\delta_{\rm p}$)
%between the Li plane and the vicinity of O$^{2-}$ ions
%is about 2~eV for Li$_{0.75}$CoO$_2$.
%Such $\delta_{\rm p}$ seems too large compared to
%the difference between $T_d^{\rm Li}$ and $T_d^{\mu}$ ($\delta T_d\sim150~$K),
%even if we consider the mass difference.
%However, since the diffusion is dominated
%by the potential barrier between the nearest sites,
%it would be reasonable that $\delta_{\rm p}\gg\delta T_d$.

\begin{figure}[t]
  \begin{center}
    \includegraphics[keepaspectratio=true,width=60 mm]{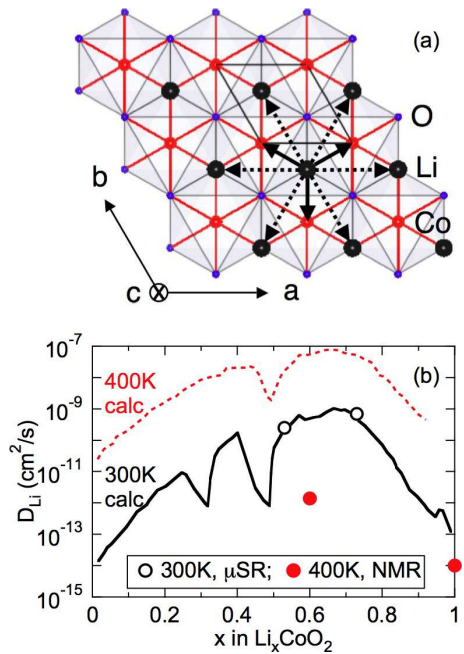}
  \end{center}
  \caption{(Color online)
(a) Possible jump paths for Li ions.
Broken arrows represent the direct jump to nearest (vacant) Li site
(path No.~1), whereas solid arrows the jump to an interstitial site 
in the center of the oxygen tetrahedron
(path No.~2).
(b) The relationship between $D_{\rm Li}$ and $x$ in Li$_x$CoO$_2$ 
as extracted from our $\mu^+$SR experiment (open circles). 
Solid and dashed lines represent the predictions by first-principles calculations
\cite{Ven} at $T$~=~300 and 400~K, respectively, 
when the effective vibration frequency is a typical value (10$^{13}$~s$^{-1}$).
Sharp minima in the predicted curve (at $x=1/3$ and 1/2) are caused by Li-ordering.
The $^7$Li-NMR result \cite{Nakamura_2,Nakamura_3} 
for $T$~=~400~K is also plotted (solid dot) for comparison.
}
  \label{fig:DLi}
\end{figure}

Finally, we estimate $D_{\rm Li}$ using the obtained fluctuation rate $\nu$ as
directly measuring the jump rate.
Figure~\ref{fig:DLi}(a) shows the possible jump paths for the Li ions.
That is, the direct jump to the nearest (vacant) Li site
(path No.~1) and
the jump to the interstitial site in the center of the oxygen tetrahedron
(path No.~2).
Assuming that $\nu$ corresponds to the jump rate of the Li ions
between the neighboring sites,
$D_{\rm Li}$ is given by \cite{Borg};
\begin{eqnarray}
D_{\rm Li}&=&
\sum^{n}_{i=1}\frac{1}{N_i}Z_{v, i}s_i^2\nu,
\label{eq:DLi}
\end{eqnarray}
where $N_i$ is the number of Li sites in the $i$-th path,
$Z_{v, i}$ is the vacancy fraction, and
$s_i$ is the jump distance.
Here, we naturally restrict the path to lie in the $c$-plane,
i.e., along the 2D channel,
because it is most unlikely that the Li ions jump across the CoO$_{2}$
plane to an adjacent Li plane.
%because it is most unlikely to jump to the Li site in the adjacent Li plane
%across the CoO$_2$ plane.
Therefore, $N_1=6$, $N_2=3$,
$s_1$ is equivalent to the $a$-axis length,
$s_2=a/\sqrt{3}$,
$Z_{v, 1}=0.27$ for Li$_{0.73}$CoO$_2$ (0.47 for Li$_{0.53}$CoO$_2$),
and $Z_{v, 2}=1$. 
As a result, we obtain $D_{\rm Li}=(7\pm2)\times10^{-10}$~cm$^2$/s 
[$(2.5\pm0.8)\times10^{-10}$~cm$^2$/s]
for Li$_{0.73}$CoO$_2$ [Li$_{0.53}$CoO$_2$] at 300~K. 
Here, 
$\nu(300~$K) for Li$_{0.73}$CoO$_2$ was estimated from the extrapolation 
of the linear relationship between log[$\nu$] and $T^{-1}$ 
[see Fig.~\ref{fig:DN}(g)].
The estimated $D_{\rm Li}$ is found to be very consistent
with the prediction by first-principle calculations \cite{Ven},
as seen in Fig.~\ref{fig:DLi}(b). 
Note that the jump paths used in Eq.~(\ref{eq:DLi}) are 
the same to those for the first-principle calculations. 
This means that there is no ambiguous factor 
for estimating $D_{\rm Li}$ by $\mu^+$SR. 
Since $\mu^+$SR detects $\nu$ 
ranging from $\sim0.01\Delta$ to $\sim10\Delta$, 
it is applicable for materials with $D_{\rm Li}=10^{-12}-10^{-9}~$cm$^2$/s, 
when $N=10$, $Z_{v}=1$, $s=1~$nm, and $\Delta=0.1\times10^6~$s$^{-1}$.

%{\color {red} (Referee B, $\sharp1$)
%Here, we wish to address past $\mu^+$SR work on the LiMn$_2$O$_4$ spinel 
%\cite{Kaiser,Ariza},   
%in which $D_{\rm Li}$ was not clearly determined from its KT parameters.  
%According to dipole field calculations, 
%the muon sites are most likely to be the same to the Li site in the spinel lattice. 
%Furthermore, the KT parameters of LiMn$_2$O$_4$ are strongly affected 
%by a Jahn-Teller transition around 285~K \cite{Jun_3}.  
%As a result, it would be difficult to estimate $D_{\rm Li}$ directly from $\nu$ for LiMn$_2$O$_4$, 
%in contrast to the present Li$_x$CoO$_2$. 
%}

In conclusion, we have been able to determine the Li diffusion coefficient, 
$D_{\rm Li}$, of Li$_{x}$CoO$_2$  from the 
fluctuation rate of the field experienced by the muons in interaction with
the nuclear moments of the diffusing ions.   
%as extracted from a $\mu^+$SR experiment. 
The value was found to be in 
good agreement with theoretical predictions.
Consequently, we would like to suggest $\mu^+$SR as a novel probe to
investigate Li diffusion, especially for materials containing transition metal ions. 

This work was performed at the RIKEN-RAL Muon Facility at ISIS, and 
we thank the staff for help with the $\mu^+$SR experiments.
We appreciate T. Ohzuku and K. Ariyoshi for sample preparation and 
K. Yoshimura for discussion.
JS and YI are supported
by the KEK-MSL Inter-University Program for Overseas Muon Facilities.
This work is also supported by Grant-in-Aid for Scientific Research (B),
19340107, MEXT, Japan.
The image involving crystal structure was made with VESTA.

\end{document}